\documentclass[onecolumn,showpacs,amsmath,amssymb,prd,nofootinbib]{revtex4-2}
\usepackage{graphicx}
\usepackage{epsfig}
\usepackage{enumerate}
\def\beq{\begin{equation}}
\def\eeqno#1{\label{#1}\end{equation}}

\def\rarrow{\rightarrow }

\def\dleft{\rlap{{\it D}}\raise 8pt
\hbox{$\scriptscriptstyle\Leftarrow$}}
\def\dright{\rlap{{\it
D}}\raise 8pt\hbox{$\scriptscriptstyle\Rightarrow$}}

\def\az{a_{0}}

\def\l0{\ell_{0}}

\def\rar{\rightarrow}
\def\s{\sigma}

\def\a{\alpha}
\def\b{\beta}
\def\c{\gamma}
\def\l{\lambda}

\def\f{\phi}
\def\t{\theta}

\def\k{\kappa}
\def\r{\rho}
\def\m{\mu}

\def\z{\zeta}

\def\av#1{\langle#1\rangle}

\def\A{\mathcal{A}}

\def\F{\mathcal{F}}
\def\L{\mathcal{L}}
\def\O{\mathcal{O}}

\def\EE{\mathcal{E}}

\def\o{\omega}

\def\D{\Delta}
\def\Lap#1{\D^{[#1]}}

\def\d{\delta}

\def\a{\alpha}
\def\xlimin{{x\rarrow\infty \atop{\raise 1pt\hbox to 30pt
{\rightarrowfill}}}}
\def\limlim#1#2{{#1\rarrow #2 \atop{\raise 1pt\hbox to 30pt
{\rightarrowfill}}}}
\def\eps{\epsilon}
\def\vr{{\bf r}}

\def\vX{{\bf X}}

\def\vR{{\bf R}}
\def\vP{{\bf P}}
\def\vF{{\bf F}}
\def\vv{{\bf v}}

\def\vg{{\bf g}}
\def\vk{{\bf k}}

\def\va{{\bf a}}

\def\vF{{\bf F}}

\def\S{\Sigma}

\def\grad{\vec\nabla}
\def\div{\vec \nabla\cdot}
\def\gf{\grad\phi}

\def\L{\mathcal{L}}

\def\G{\mathcal{G}}
\def\fN{\f\_N}
\def\gN{g\_N}

\def\M{\mathcal{M}}
\def\m{\mu}
\def\a{\alpha}
\def\b{\beta}

\def\F{\mathcal{F}}

\def\D{\Delta}

\def\_#1{_{\scriptscriptstyle #1}}
\def\^#1{^{\scriptscriptstyle #1}}

\def\azg{\A_0}

\def\oot{\frac{1}{2}}

\def\CC{\mathcal{C}}

\def\GG{{\vec\G}}
\def\CCv{{\vec\CC}}
\def\vgN{\vg\_N}

\def\rro#1{\overset{\text{\scriptsize$\leftrightarrow$}}{#1}}
\def\rM{r\_M}
\begin{document}
%\eqsec  % uncomment this line to get equations numbered by (sec.num)
\title{Is MOND necessarily nonlinear?}

\author{Mordehai Milgrom}
\affiliation{Department of Particle Physics and Astrophysics, Weizmann Institute}

\begin{abstract}

The iconic, deep-MOND-limit (DML) relation between acceleration and mass, $a\sim (M\azg)^{1/2}/r$, implies that, in MOND, {\it accelerations} cannot be linear in the mass distribution ($\azg\equiv G\az$ is the DML constant, and $\az$ the MOND acceleration). This leads to important idiosyncracies of MOND, such as a breakdown of the strong equivalence principle, and the resulting ``external-field effect'' (EFE).
I show that the DML axioms are, in themselves, consistent with a, possibly unique, nonrelativistic, action-based, {\it linear} formulation of the DML: For a test particle in the gravitational field of a mass distribution $\r(\vr)$, the equation of motion is $d^4\vr/dt^4=\grad\psi$, where $\psi(\vr)=-\frac{\azg}{2}\int\frac{\r(\vr') d^3\vr'}{|\vr-\vr'|^2}$. This model suffers from important drawbacks, which may make it unacceptable as a basis for a full-fledged MOND theory, which may imply that the answer to the title question is ``yes.'' The model is unique among MOND theories propounded to date not only in being linear -- hence not exhibiting an EFE, for example -- but in constituting a modification of both Newtonian inertia and Newtonian gravity. This linear and time-local model inspires and begets several one-parameter families of models. One family employs nonlinear, time-nonlocal kinetic terms, but still linear gravitational-field equations.  Other families generalize the DMLs of AQUAL and QUMOND, modifying gravity as well as inertia.
All families employ fractional time derivatives and possibly fractional Laplacians. While, at present, I cannot base some acceptable MOND theory on these models -- for example, I cannot offer a sensible umbrella theory that interpolates between these DML models and Newtonian dynamics -- they are useful in elucidating various fundamental aspects of MOND. For example, they help to understand which predictions follow from only the basic tenets of MOND -- so-called primary predictions -- and which are secondary, i.e., theory dependent. For example, they demonstrate how the EFE can be strongly theory dependent, not only in its strength but in its very nature. The models may also help show the way to a wider class of MOND theories. Finally, I describe another family of models that are all linear, but non-MONDian, having Newtonian dynamics at one end and the linear DML model at the other.

\end{abstract}
%\keywords{}
%\pacs{04.50.Kd, 98.80.Jk}
\maketitle

\section{Introduction   \label{introduction}}
The MOND paradigm \cite{milgrom83} accounts for the mass anomalies in galactic systems, without dark matter, by modifying Newtonian dynamics and general relativity in the limit of small accelerations, as are found in such systems. It introduces a new constant$, \az$, with the dimensions of acceleration, at or below which dynamics depart from the old dynamics. In the ``deep-MOND''  limit (DML) -- much below $\az$ -- MOND dictates that dynamics of systems governed by gravity become spacetime scale invariant -- i.e., invariant under $(\vr,t)\rar\l(\vr,t)$ -- at least in the nonrelativistic limit \cite{milgrom09}. In the high-acceleration limit -- much above $\az$ -- MOND posits a rapid approach to standard dynamics -- a ``correspondence principle.'' Reviews of MOND can be found in Refs. \cite{fm12,milgrom14,milgrom20a,mcgaugh20,merritt20,bz22,fd25}.
\par
Newtonian gravitational dynamics is linear, since the acceleration at position $\vr$ is a linear functional of the mass density, $\r$. Specifically,
\beq   \frac{d^2\vr}{dt^2}=\vgN(\vr)=\int d^3\vr' \r(\vr')\vec\G(\vr-\vr');~~~~~\vec\G(\vr)\equiv -G\frac{
 \vr}{|\vr|^3}.\eeqno{newt}
\par
In contradistinction, the emblematic deep-MOND relation between the mass distribution that sources gravity, and the acceleration it imparts to bodies, $a\approx (\az \gN)^{1/2}$ -- where $\gN$ is the Newtonian gravitational acceleration that scales as $M/r$ -- tells us that the DML, and hence MOND at large, cannot be linear in the same sense as Newtonian gravitational dynamics.
\par
However, a linear relation between accelerations and $\rho$ is not the general expression of linearity.
\par
More generally, in the context of nonrelativistic gravity, which I concentrate on here, strict linearity can be taken to mean that the trajectory, $\vr(t)$, of a test particle is determined by an equation of motion (hereafter EoM) of the form
\beq (\vec\O\vr)(t)=(\CCv\r)(\vr), \eeqno{wusda}
where $\vec\O$ is a linear, possibly time-nonlocal, operator acting on the trajectory $\{\vr\}$ to give another vector function of time, ($\vec\O\vr)(t)$, and $\CCv$ is a vector, linear operator acting on the field $\r$.\footnote{When $\vec\O$ is time nonlocal, $\{\vr\}$ is the full trajectory, not the local value of $\vr$, and a finite number of its time derivatives, at time $t$.}
\par
The particle's mass is assumed not to appear in the EoM so as to ensure the universality of free fall (independence of the trajectory on the mass of the test particle).
\par
We ask whether there is a DML theory that is linear in the more general sense of Eq. (\ref{wusda}).
\par
Note that special relativity is not linear even in this more general sense. Its EoM for a particle is of the form
\beq \frac{md(\c\vv)}{dt}\equiv m\rro{\m}\left(\frac{\vv}{c}\right)\va=\vF.  \eeqno{mayata}
Here, $\vF$ is the force acting on the particle, $\c$ is the particle's Lorentz factor, and the ``extrapolation matrix'',
$\rro{\m}$ is
\beq \rro{\m}\left(\frac{\vv}{c}\right)\equiv\c (1+\c^2\frac{\vv\otimes\vv}{c^2}). \eeqno{damure}
In the Newtonian limit, $c\rar\infty$, $\rro{\m}$ tend to the unit matrix.
For example, for a charge in an electromagnetic field, $\vF$ is the Lorentz force, which is linear in the sources (charges and currents).
But the left-hand side is not linear in the trajectory.
\par
Special relativity is, nonetheless, ``quasilinear'' (to use the standard mathematical nomenclature) in that the left-hand side of the EoM is linear in the highest derivative of the trajectory (the second in this case) and the nonlinearity enters only through the lower derivatives (the first in this case).\footnote{Note that this is not the sense in which MOND theories, such as QUMOND and its generalizations, are ``quasilinear''.}
\par
Thus, in principle, one should consider, more generally, also MOND theories whose EoM for a particle trajectory is of the form
\beq \EE[\vv,\va,\vr^{(3)},...,\vr^{(n-1)}]+\F[\vv,\va,\vr^{(3)},...,\vr^{(n-1)}]\vr^{(n)}=(\CCv\r)(\vr) \eeqno{gurda}
[$\vv=\dot\vr,~~\va=\ddot\vr$].
If the kinetic, left-hand side is derivable from a time-local Lagrangian, $L[\vv,\va,\vr^{(3)},...,\vr^{(k)}]$, then it is quasilinear, since the highest derivative (of order $2k$) in the Euler-Lagrange equation appears only in a term of the form
\beq  \frac{\partial^2L}{\partial\vr^{(k)}\partial\vr^{(k)}}\cdot \vr^{(2k)}.\eeqno{highest}
\par
We use such a time-local EoM -- one that involves a finite number of derivatives -- to propagate a trajectory in time from some initial conditions. We know all the lower-than-highest derivatives at any given time step, and we use the equation to calculate the highest one, which then does become a linear functional of the mass distribution. However, MOND theories that modify Newtonian inertia are, generically, time nonlocal. This is implied by a ``theorem'' I proved in Ref. \cite{milgrom94a}, to the effect that a MOND theory that modifies inertia, that has the standard symmetries, {\it including Galilei invariance}, and that has the correct Newtonian and deep-MOND limits, must be time nonlocal. In such nonlocal theories we do not propagate a solution in time, but solve for trajectories with their entire time dependence at once. So, quasilinearity is not relevant in this case.
\par
Here I investigate whether and to what extent there exist theories that are compatible with the axioms of MOND and yet are linear in the strict sense of Eq. (\ref{wusda}). In doing this, I greatly expand on some arguments addressing this question, discussed in Ref. \cite{milgrom14a}.
\par
Since the aim here is heuristic, I will concentrate on the easier task of studying only pure DML theories, without worrying whether there is an umbrella theory that naturally interpolates them with a Newtonian limit as accelerations sweep across $\az$.
This frees us from a crucial requirement in the above theorem, and may allow construction of theories that are local and thus are much more wieldy.\footnote{The recently propounded theory in Ref. \cite{blanchet25} is an example of a pure DML theory, still lacking a high-acceleration limit that gives standard dynamics.}
\par
In Sec. \ref{path}, I show how the DML axioms greatly constrain linear DML theories, leading to practically a single such model theory.
In Sec. \ref{predictions}, I derive some predictions of this model, showing how it, naturally, shares the primary MOND predictions, but can differ greatly from other MOND theories on secondary predictions.
Section \ref{limitations} exposes some of the problems and limitations of the linear model, which renders it, at present, unsuitable as a full fledged theory, but still leave it a useful heuristic tool. Section \ref{generalizations} describes a generalization to a one-parameter family of DML models, with inertia that is nonlinear and nonlocal, but with a gravitational sector that is still linear. I also discuss some predictions of such models. These models are free of some of the problems of the linear model. Section \ref{shell} describes two model families that generalize the DMLs of AQUAL and QUMOND, respectively. In Sec. \ref{family}, I describe another one-parameter family of models that span the gamut between the linear DML model and Newtonian dynamics.
Section \ref{lessons} summarizes some of the lessons we can learn from the preceding discussions.

\section{Narrow path to linear deep-MOND \label{path}}
The generic nonlinearity of MOND enters through the DML. So, I start by considering whether it is possible to write a linear theory for the pure DML. Thus, consider the form that the linear equation (\ref{wusda}) has to take, assuming only the MOND basic tenets that concern the DML. Namely, we require that only $G$ and $\az$ appear in the equations, and that they only appear in the combination $\azg=G\az$ -- a requirement that is equivalent to requiring scale invariance.\footnote{This was shown, e.g., in Ref. \cite{milgrom14a}, and it requires that we normalize all degrees of freedom so that their scaling dimension agrees with their $[m][\ell][t]$ dimensions -- which is always possible by multiplying them by powers of $\az$ and $G$. For example, The Newtonian gravitational potential, which scales as $MG/r$, transform under scaling as $\fN\rar\l^{-1}\fN$ which does not agree with its dimensions being $[\ell]^2[t]^{-2}$. We should then work with $\bar\fN=\az\fN$, which does have matching dimensions.} In addition we assume that the action/EoM does not have explicit time dependence -- otherwise, we would have to introduce an additional constant with the dimensions of time. We also assume, as a matter of course, translation and rotation invariance.
\par
It is easy to see that under these requirements, linearity is not possible to implement in theories that can be described as pure modified gravity.
In (nonrelativistic) modified-gravity MOND theories of the general class discussed in Ref. \cite{milgrom14b} (AQUAL \cite{bm84}, QUMOND \cite{milgrom10}, and TRIMOND \cite{milgrom23b} are special cases), the EoM is still of Newton's-second-law form $\va=-\gf$, but the DML potential is determined from $\r(\vr)$ by some modified DML prescription, not by the Poisson equation. Whatever the prescription is, dimensional arguments show that $\gf$ cannot be a linear functional of $\r$, but that it has to scale as $\r^{1/2}$: To see this, note that, in such theories, the left-hand side of the EoM (i.e., $\va$) has dimensions $[\ell][t]^{-2}$. The right-hand side is constructed from $\azg$ (of dimensions $[m]^{-1}[\ell]^4[t]^{-4}$), $\r$ (of dimensions $[m][\ell]^{-3}$), and $\vr$. If we do not allow explicit time dependence, the only possibility to combine these to get the right dimensions must scale as $(\azg\r r)^{1/2}$, which scales as $(\az GM/r^2)^{1/2}\sim (\az a\_N)^{1/2}$. Hence, nonlinearity follows inevitably.\footnote{These expressions are, of course, not algebraic, local relations between the DML acceleration $\gf$ and the density; they only give us the dimensional scaling of the former with the latter. For example, in AQUAL, $\gf$ is determined from the field equation $\div(|\gf|\gf)\propto \azg\r$, and in QUMOND it is determined from $\D\f\propto \az^{1/2}\div(|\gf\_N|^{-1/2}\gf\_N)$, where $\gf\_N$ is determined from $\div\gf\_N\propto G\r$.}
\par
Thus, if we want to seek linear, DML EoMs, we must seek them among so-called modified-inertia theories, those that modify the inertial term and possibly also the gravitational one.
\par
In Ref. \cite{milgrom94a}, I showed that a MOND theory (hinging only on $\az$ and $G$) derivable from a Lagrangian, that modifies the kinetic Lagrangian, and that has time-translation, space-translation, rotation, and Galilei invariance, and that has the correct MOND and Newtonian limits, must be time-nonlocal. Namely, its Lagrangian cannot be a function of only a finite number of time derivative, or even a limit of a series of such Lagrangians.
However, with all its nonlocality, the Lagrangian of such a theory does have become local, of course, in the Newtonian limit, achieved formally by taking the limit $\az\rar 0$. Likewise, it is possible that such a theory also becomes local in the very deep-MOND limit (DML), achieved formally in the limit $\az\rar\infty$ and $G\rar 0$, while $\azg\equiv \az G$ remains constant \cite{milgrom14a}.
Indeed, the DML model I shall construct is local and modifies inertia (as well as gravity). So the theorem will tell us that it does not have a local umbrella theory that interpolates it with Newtonian dynamics. If there is one it will be nonlocal.
\par
Thus, we want to construct a DML EoM for the trajectory $\vr(t)$ of a test particle of mass $m$ under the gravitational influence of a (bounded) mass distribution $\r(\vr)$.
Among the system attributes and constants only $\r$, $m$, and $\azg$ have mass dimensions ($m^{-1}$ in $\azg$). Since we want to retain the universality of free fall, $m$ cannot appear.
So, Eq. (\ref{wusda}) must have the form
\beq (\vec\O\vr)(t)=\azg(\GG\r)(\vr), \eeqno{wusda1}
where $\GG$ is a linear,  vector operator with only time and length dimensions, and $\azg$ not appearing anywhere else.
If we apply the model to stationary density distributions, and we do not have dimensioned constants to use in the construction, $\GG$ must not have time dimensions.
Then, the right-hand side, and hence the left-hand one, has time dimensions $[t]^{-4}$ (from $\azg$).
Since we want to maintain the linearity of the left-hand side in $\vr$, it must have dimensions of $[\ell][t]^{-4}$. This implies that $\GG$ is dimensionless.
\par
Time-translation invariance, and the absence at our disposal of a constant with the dimensions of time, dictate that
 $\vec\O\vr\propto d^4/dt^4$.
If we further impose translational and rotational invariance, a simple choice of $\GG$ gives the EoM
 \beq \frac{d^4\vr}{dt^4}=\azg\int d^3r'\frac{\r(\vr')
 (\vr-\vr')}{|\vr-\vr'|^4}=\grad\psi(\vr), ~~~~~~~~\psi(\vr)=-\frac{\azg}{2}\int d^3r'\frac{\r(\vr')}{|\vr-\vr'|^2}.  \eeqno{pasafal}
\par
This is the result arrived at, and discussed briefly in Ref. \cite{milgrom14a}. Without having a proof to this effect, this model seems unique given our requirements.\footnote{For example, expressions for $\psi$ that involve derivative of $\r$ are equivalent for bounded mass distribution, as can be shown by integration by parts.} But it would be interesting to explore this issue further.
 \par
The sole appearance of $\azg$ ensures that for any physical trajectory $\vr(t)$ around a point mass, the trajectory $\vr_\a(t)\equiv \a\vr(t/\a)$ is also a physical trajectory \cite{milgrom09}.\footnote{While this follows straightforwardly from the specific form of the EoM, this result holds in any gravitational theory, where the central mass determines the dynamics, and which involves only $\azg$ as a dimensioned constant.} For circular trajectories, it follows that the circular speed, $V_c$, around a point mass, $M$, is independent of the orbital radius; it also follows that $V_c\propto(M\azg)^{1/4}$.
The normalization of the right-hand side of the EoM is chosen according to the standard normalization of $\az$, which corresponds to equality $V_c=(M\azg)^{1/4}$. The last two results are primary MOND predictions and follow from only the basic tenets (and the additional, small-print assumption that dynamics of a test particle far from a bounded system depend only on the total mass, $M$).
\par
For a system of point masses, $m_p$, with trajectories $\vr_p(t)$ -- which I will consider hereafter -- the EoM (\ref{pasafal}) gives
\beq \frac{d^4\vr_p}{dt^4}=\azg\sum_{q\not = p}\frac{m_q
(\vr_p-\vr_q)}{|\vr_p-\vr_q|^4}=\grad\psi(\vr_p),  \eeqno{pasafma}
with
\beq \psi(\vr)=-\frac{\azg}{2}\sum_{q\not = p}\frac{m_q}{|\vr-\vr_q|^2}.  \eeqno{pasafalma}
\par
This model can be derived from the action $I\propto \int dt~L(\vr_p,\ddot\vr_p)$, with the Lagrangian
\beq L=\sum_p \oot m_p\ddot\vr_p^2+\oot\azg\sum_{q<p}\frac{m_pm_q}{|\vr_p-\vr_q|^2}.    \eeqno{lag}
\par
In the formulation above, $\psi$ is only an auxiliary field; it is not treated as a system degree of freedom, with its own ``free'' term in the Lagrangian.
As is done in going from the Newtonian expression of the potential in terms of the density to the Poissonian formulation of Newtonian dynamics, here too, we can introduce $\psi$ as a gravitational degree of freedom and start from the Lagrangian (for the continuum description)
\beq L= \int d^3r\{\azg\r[\oot(\ddot\vr)^2-\psi]+\F(\psi)\}.  \eeqno{genlagut}
$\F(\psi)$ is a functional of $\psi$. This form implies that $\psi$ has dimensions of acceleration squared, and its scaling dimension must be $-2$ (namely, under scaling $\psi\rar\l^{-2}\psi$). Varying over the matter position degree of freedom, we get
\beq \frac{d^4\vr_p}{dt^4}=\grad\psi(\vr_p),  \eeqno{pasalul}
as in Eq. (\ref{pasafal}).
We want to take $\F$ so that all terms in $L$ transform in the same way under scaling -- so as to ensure scale invariance of the EoMs.
In the present context we also want $\F$ to be quadratic in $\psi$, so that we get a linear theory. Such dimensional considerations imply, straightforwardly, that $\F(\psi)$ scales as $\psi^2/r$. I also require that $\psi$ itself does not appear in $\F$, only its derivatives, so that the theory is invariant under $\psi\rar\psi+ {\rm const.}$ The only choice I can think of (up to divergences, which do not change the field equation) is $\F(\psi)=-q(\Lap{1/4}\psi)^2$, with the coefficient $q$ to be chosen in alignment with the standard normalization of $\azg$. Anticipating subsequent derivation, we take $q=1/2\pi^2$.
We thus end up with the Lagrangian
\beq L= \int d^3r\left\{\azg\r[\oot(\ddot\vr)^2-\psi]-\frac{1}{2\pi^2}(\Lap{1/4}\psi)^2\right\}.  \eeqno{genlag}
(Note that the Lagrangians used in this paper may differ by which of the terms take the prefactors involving the constants, $G,~\az,~\azg$, according to convenience. This has no effect on the resulting theory because going from one form to another involves multiplying the whole action by a constant.)
\par
The expression for the Lagrangian (\ref{genlag}) employs a fractional-Laplacian operator, $\Lap{1/4}$, which is, in a sense, a spatial derivative of order $1/2$. More generally, I use $\Lap{\a}$, here and elsewhere in this paper, whose exact definition and manipulations are described in Appendix \ref{fractional}.\footnote{There are various ways to do this. The subject of fractional calculus has a long history. More recent accounts can be found, e.g., in Refs.
\cite{caputo67,hilfer00,herrmann14,kwasnicki17,garofalo17,stinga19,giusti20}.} According to this definition,  $\Lap{1}=\D$ is the standard Laplacian.
\par
I show in Appendix \ref{fractional} that in varying over $\psi$
\beq \d\int d^3r(\Lap\a\psi)^2=-2\int d^3r(\Lap{2\a}\psi)\d\psi.  \eeqno{maca}
Varying the action over $\psi$ we thus get the field equation
\beq \Lap{1/2}\psi=\pi^2\azg\r.  \eeqno{fieldps}
I show in Appendix \ref{fractional} that the Green's function of $\Lap{1/2}$ is
\beq \G_{1/2}(\vr)=-\frac{1}{2\pi^2\vr^2};   \eeqno{greenhalf}
namely,
\beq \Lap{1/2}\G_{1/2}(\vr)=\d^3(\vr),   \eeqno{greenba}
with the appropriate boundary conditions on $\psi$.
From this follows that $\psi$ is given by expression (\ref{pasafal}), which also justifies our choice of $q$.
\par
As in the Newtonian case, such a formulation with $\psi$ a degree of freedom, satisfying a {\it linear} field equation, can be very useful, as it lands itself readily to solution methods based on Fourier transform.
\par
The above very restricted form, forced on us by the DML basic tenets, and the symmetry requirements, involves a combined modification of Newtonian gravity (since the Newtonian gravitational potential does not play any role), and of Newtonian inertia (as Newton's second law does not apply).
{\it It is the fact that $\azg$ has mixed length and time dimensions that forces us to modify both the inertial and the gravitational terms in order to get a theory that is both linear and scale-invariant.}
\par
If we introduce, instead, a new length constant, $\ell_0$, for example, we can construct a scale-invariant, linear EoM by modifying only the gravitational part. This is done, e.g., in Ref. \cite{giusti20}, where one replaces $\vec\G$ in the right-hand side of the Newtonian EoM (\ref{newt}) with $\vec\G=-G\vr/r^2\ell_0$, or, equivalently, replace the Poisson equation for the gravitational potential by one involving a fractional Laplacian: $\D\f\propto \r~\Rightarrow~\ell_0\Lap{3/2}\f\propto \r$.
Such models predict asymptotically flat rotation curves -- which follows only from the scale invariance. However, they predict the wrong mass asymptotic speed relation; namely, they predict $V^2_\infty\propto MG/\ell_0$, instead of the observed $V^4_\infty\propto M$. To repair this, and reproduce the prediction of MOND, the author of Ref. \cite{giusti20} and others who attempt to reproduce MOND by introducing a scale length, make the ansatz that $\ell_0$ is not a constant of the theory, but instead $\ell_0\propto \rM=(MG/\az)^{1/2}$, where $\rM$ is the MOND radius of the total mass of the body in question. I think that this is unacceptable if only because ``the total mass'' is not defined since every system (e.g., a galaxy) is part of larger and larger systems. Which ``total mass'' is then to enter the gravitational field at some specific point in the Universe?
\par
Similarly, introducing a frequency constant $\o_0$  enables us to construct  a linear, scale-invariant theory by modifying only the inertial term.

\section{Some generic predictions of the linear DML model and comparison with other MOND theories  \label{predictions}}
Here I consider the linear, DML model arrived at in Sec. \ref{path} in some more detail. I discuss some of its consequences, distinguishing primary DML predictions from secondary ones, and comparing with predictions of other MOND theories. Some MOND predictions concern the transition from the DML to the Newtonian domain -- such as the onset of the mass discrepancies always occurring at accelerations around $\az$. Clearly, such predictions are beyond the power of this model, which is a strict DML one, without transition to the Newtonian regime.
Predictions that can be studied in the model concern, e.g., asymptotic behavior of rotation curves, the mass-asymptotic-speed relation, the relation between the velocity dispersion in a pressure-supported system and the system's total mass, the center-of-mass motion of a composite system, and whether it is dictated by the kinematics of that center, or is affected by the kinematics of the constituents, and the workings of an external-field effect (EFE).
\par
Note that the general, two-parameter scaling of the DML holds here:
If $\vr_p(t)$ is a system history for masses $m_p$, then  $\l\vr_p(t/\k)$ [with velocities $(\l/\k){\bf v}_p(t/\k)$], is a system history, for masses $(\l/\k)^4 m_p$ for any $\l,\k>0$. For continuum systems, if $\rho(\vr,t)$, $\vv(\vr,t)$ is a solution, so is $\l\k^{-4}\rho(\vr/\l,t/\k)$, $(\l/\k)\vv({\bf r}/\l,t/\k)$. (This is the scaling that, when applied to $[\ell,~t,~m]$ units, does not change the value of $\azg$.)
\subsection{Center-of-mass motion \label{comm}}
Consider a system of point masses, $m_p$, bounded within a region of radius $a$, together with masses, $\M_s$, external to the system, and at distances $L_s\gg a$. All masses move, individually, according to our linear model. Then, the equation of motion of the system's center of mass, $\bar\vR$ due to $\M_s$ is also described correctly by the model, if we take the system to be a point mass, $M=\sum_p m_p$, irrespective of the internal dynamics of the system itself. To see this, take the origin inside the system, and the radius vectors of $M_s$ as $\vR_s$, with $|\vR_s|\gg |\vr_p|$ for all $p$ and $s$. We then have
 \beq \frac{d^4\vr_p}{dt^4}=\azg\sum_{q\not = p}\frac{m_q
 (\vr_p-\vr_q)}{|\vr_p-\vr_q|^4}+\azg\sum_s\frac{\M_s
  (\vr_p-\vR_s)}{|\vr_p-\vR_s|^4}.  \eeqno{paster}
 Multiplying by $m_p$, summing, and replacing the denominator in the second term by $|\bar\vR-\vR_s|^4$ (justified because $|\vr_p|,~|\bar\vR|\ll |\vR_s|)$, the sum over the first term drops, and we get in the limit
 \beq \frac{d^4\bar\vR}{dt^4}=\azg\sum_s\frac{\M_s
 (\bar\vR-\vR_s)}{|\bar\vR-\vR_s|^4},  \eeqno{pasterma}
 which justifies the above statement. This, of course is the result of the linearity of the model.
 \par
It follows that in a system of well-separated (many-body) subsystems -- subsystems much smaller than their separations -- we can represent the subsystems by point masses (irrespective of their internal dynamics) if we only want to describe the center-of-mass motions of the subsystems.
 We can then write the virial relation for only the relative motions.
 \par
Since $\bar\vR=\sum_p m_p\vr_p$, satisfies $d^4\bar\vR/dt^4=0$ for an isolated system, we identify the conserved momentum as
 \beq \bar\vP=\pm M d^3\bar\vR/dt^3. \eeqno{moma}
 This follows also from the general Ostrogradsky, Hamiltonian formalism for higher-derivative theories (\cite{woodard15}, and see Sec. \ref{hamiltonian} below), where it follows formally that the particle momentum should be identified with $-md^3\vr/dt^3$.
 \par
 Multiplying the EoM (\ref{pasafma}) by $\dot\vr_p$ shows that for a particle in static field $\psi(\vr)$, the particle attribute $E_p=E_{Kp}+m_p\psi(\vr_p)$ is conserved, with the particle kinetic energy
 \beq   E_K\equiv m[\frac{(\va)^2}{2}-\dot\va\cdot\vv]    \eeqno{energ}
  ($\vv=\dot\vr$ is the velocity, and $\va=\ddot\vr$ is the acceleration).
 For an isolated many-body system, the total energy $\sum_p [E_{Kp}+m_p\psi(\vr_p)]$ is conserved.
 This definition of the energy also follows from the Hamiltonian formalism.

\subsection{Virial relation   \label{virial}}
Take the scalar product of Eq. (\ref{pasafma}) with $m_p\vr_p$ and sum over $p$ to get
 \beq \frac{d}{dt}[\sum_p m_p (\vr_p\cdot\dot\va_p-\vv_p\cdot\va_p)]+\sum_p m_p \va_p^2=\azg\sum_{q< p}\frac{m_p m_q}{|\vr_p-\vr_q|^2}.  \eeqno{vir1}
In quasistatic systems, in which the macroscopic attributes (such as the sum in the first term) are time independent, the first term vanishes, and we have
\beq \av{\va^2}=M\azg\sum_{q< p}\frac{\m_p \m_q}{|\vr_p-\vr_q|^2},  \eeqno{vir2}
where, $\av{\va^2}=M^{-1}\sum_p m_p \va_p^2$ is the mass-weighted average of the squared acceleration, $M$ is the total mass of the system, and $\m_p=m_p/M$ are the mass fractions of the constituents.\footnote{If the system is not quasistatic, but still bounded, the long-time average of the first term in Eq. (\ref{vir2}) vanishes, and the relation is understood as its long-time average.}$^,$\footnote{As a general comments on such virial relations: If the system is made up of subsystems that are small compared with their interdistances -- such as a galaxy made of stars -- it may not be useful to calculate the contribution to the quantities in the relation from the interiors of the subsystems. Instead, it is justified to replace the subsystems by point masses, whose dynamics are accounted for by the virial relation.}
 \par
 Relation (\ref{vir2}) is to be compared with the DML virial relation that holds in a large class of modified-gravity theories \cite{milgrom14b}
 \beq \av{V^2}^2=\frac{4}{9}M\azg[1-\sum_p \m_p^{3/2}]^2,\eeqno{virimg}
 where $\av{V^2}$ is the mass-weighted mean of the square of the constituent velocities, in a frame where the system is static (the derivation assumes, as above, that global properties are time independent).
 \par
 If we take as some measure of the radius of the system the quantity $R$,  defined by $(u R)^{-2}\equiv \sum_{q< p}\m_p \m_q|\vr_p-\vr_q|^{-2}$, with $u$ a numerical factor of order of a few -- $u R$ being a certain mass-weighted average of the interparticle distance -- and, defining
the system's velocity dispersion  as $\s^2=v\av{\va^2}^{1/2}R$, with $v$ some numerical geometrical factor, then we have
 a DML, mass-velocity-dispersion relation
 \beq \s^4=(v/u)^2M\azg,  \eeqno{mavel}
 which is more directly compared with relation (\ref{virimg}).
\par
As an example, for a homogeneous sphere of radius $R_0$, I find that $uR=(32/9)^{1/2}R_0\approx 1.9R_0$.
However, in general, (see Sec. \ref{limitations}), $u$ and/or $v$ can differ greatly from unity, if we take $R$ and $\s$ to be faithful representations of the system's mass and velocity dispersion. So, relation (\ref{mavel}) cannot be considered the universal $M-\s$ relation that is predicted by the basic tenets of MOND (see the explanation for this in Sec. \ref{lessons}). Relation (\ref{vir2}) is brought here only for its heuristic value, as is the linear model on which it is based.

\subsection{Spherical systems}
Since $\psi$, which determines particle dynamics, is not the Newtonian $1/r$ potential, there are no useful shell theorems to facilitate its determination for spherical configurations. $\grad\psi$ does not vanish inside a spherical hollow, and its value outside a spherical mass is not the same as that of a point mass.
\par
For example, for a thin, homogeneous, spherical shell of mass $M$ and radius $R$ we have (inside and outside)
\beq \psi(r)=-\frac{M\azg}{4Rr}\ln\left|\frac{r+R}{r-R}\right|.   \eeqno{thina}
So, $\psi(r)$ depends on $R$ for any $r$. Only asymptotically, for $r\gg R$, does $\psi\rar -M\azg /2r^2$.
We also see that $\psi$ (but not $\grad\psi$) diverges on such an ideal shell (more on this in Sec. \ref{limitations}).

\subsection{Rotation curves}
Asymptotic flatness and $V_\infty^4\propto M\az$ (equality for our normalization) follow from the DML basic tenets and are predicted in the model.
More generally, the expression for the rotational speed on circular orbits in the midplane of an axisymmetric and plane-reflection symmetric mass distribution (such as an idealized disk galaxy) is
\beq  V^4(r)=r^3\frac{\partial\psi}{\partial r}. \eeqno{rotat}
This result differs in important ways from the corresponding results in both the modified-gravity and the modified-inertia formulations considered to date. For example, the orbital acceleration,  $a=V^2/r$, is not a function of the Newtonian acceleration, $\gN$, as is the case in pure modified-inertia theories \cite{milgrom94a}, where we have $a=(\az\gN)^{1/2}$ in the DML. And, this result differs from that in a large class of modified-gravity theories, in not having shell theorems.
\subsection{The two-body problem}
Consider two bodies much smaller than their separation (``point masses'').
As a result of linearity, the two-body problem is reducible to that of a single body in a central field: If $\va_{12}=\va_1-\va_2$ is the relative acceleration of the two bodies of masses $m_1,~m_2$, then $\m\ddot\va_{12}=m_1m_2(\vr_1-\vr_2)/|\vr_1-\vr_2|^4$, where $\m= m_1m_2/(m_1+m_2)$.
\par
If the two bodies are on circular trajectories around each other, their velocity difference can be shown to be given by
\beq   V^4_{12}=M\azg,   \eeqno{twobo}
and is independent of the individual masses.\footnote{This also follows from the dynamics of the equivalent single-body system, with the expression for the reduced mass being $(m_1m_2)/M$, as in Newtonian dynamics.}
\par
This contrasts with the analogous results in modified-gravity theories, such as AQUAL and QUMOND, TRIMOND, and, in fact, the general class of modified gravity theories described in Ref. \cite{milgrom14b}:
\beq V^4_{12}= \frac{4}{9}\left(\frac{1-\m_1\^{3/2}-\m_2\^{3/2}}{\m_1\m_2}\right)^2M\azg,  \eeqno{binarmg}
where $\m_i=m_i/M$ are the fractional masses. This result also differs from that for a class of modified-inertia theories \cite{milgrom23}:
 \beq  V^4_{12}= (\m_1^{1/2}+\m_2^{1/2})^2M\azg.   \eeqno{circulla}
This demonstrates, once more, that the coefficient of $M\azg$ in the expression for $V_{12}$ is a secondary prediction, in that it is theory dependent, while the prediction that  $V_{12}/(M\azg)^{1/4}$ is of order one is a primary prediction.

\subsection{Hamiltonian formulation and the Ostrogradsky instability  \label{hamiltonian}}
Ostrogradsky \cite{ostrogradsky1850} propounded a Hamiltonian formalism for higher-derivative theories, such as our DML model, which can be very useful in analyzing such theories. A recent detailed account of the formalism is discussed in Ref. \cite{woodard15}.
For a fourth-order theory, such as the DML model derived from the Lagrangian (\ref{lag}), one defines two canonical coordinates and momenta for each particle, $p$ on a trajectory $\vr_p(t)$ \cite{woodard15} (I am omitting the particle index $p$ for clarity):
\beq \vX_1=\vr,~~~~~\vX_2=\dot\vr,~~~~~\vP_1=\frac{\partial L}{\partial\dot\vr}-\frac{d}{dt}\left(\frac{\partial L}{\partial \ddot\vr}\right),~~~~~ \vP_2= \frac{\partial L}{\partial\ddot\vr}. \eeqno{ostr}
In our case
\beq  \vP_1=-m\dot\va,~~~~~~ \vP_2=m\va.   \eeqno{caser}
The Hamiltonian is then defined as
\beq H=\sum_{s=1}^2\vP_s\cdot\vr^{(s)}-L,   \eeqno{hamilt}
which can be expressed in terms of the canonical variables as (reinstating the particle index)
\beq H=\psi(\vX_{p1})+\sum_p\frac{\vP_{p2}^2}{2m_p}+\vP_{p1}\cdot\vX_{p2}.  \eeqno{gamil}
The Hamilton equations then hold
\beq -\frac{\partial H}{\partial\vX_{ps}}=\dot\vP_{ps},~~~~~~ \frac{\partial H}{\partial\vP_{ps}}=\dot\vX_{ps},~~~~~s=1,2, \eeqno{hameq}
and encapsulate both the equations of motion and the definition of the canonical variables.
\par
The total $\vP_1=\sum_p\vP_{p1}$ is what we identified in Sec. \ref{comm} as the conserved, center-of-mass momentum.
As usual, it follows from Hamilton's equations that $H$ is conserved, i.e., $\dot H=0$.
\par
It is well documented (e.g., Ref. \cite{woodard15}) that such higher-order, local theories suffer from instabilities, owing to the fact that their Hamiltonian is not bounded from below: the last term in expression (\ref{gamil}) can be negative and arbitrarily large. The theory may then have exploding solutions.
\par
This is a serious fault in a model. However, the model is, in any event, not an acceptable description of nature, since it does not have a sensible continuation to the Newtonian regime. Such a continuation may render the model as only a limit -- not achieved in nature -- of a nonlocal (and nonlinear) theory, that, in itself, does not suffer from the Ostrogradsky instability.

\section{Limitations and deficiencies of the linear DML model   \label{limitations}}
At present, the linear DML model suffers from some deficiencies that make it unacceptable as a basis for a full fledged MOND theory. However, this model, and its generalizations to nonlinear DML models, presented below in Sec. \ref{generalizations} (and the model families described in Sec. \ref{shell}, in which the linear model is not a member) are quite useful in elucidating some fundamental issues concerning MOND.
Besides the stability issues discussed in Sec. \ref{hamiltonian}, it has the following two drawbacks.
\par
One drawback is the lack of an extension that has a Newtonian limit.
To understand why it is difficult to find an umbrella theory that has both Newtonian dynamics and the above DML model as limits, we note the following: The theorem I proved in Ref. \cite{milgrom94a} stands in the way of achieving this with a local theory -- i.e., using only derivatives of natural order all through the interpolation. More specifically, in the present case, we would have to interpolate between a $|\vr^{(2)}|^2$ DML particle kinetic Lagrangian to a  $|\vr^{(1)}|^2$ Newtonian kinetic term. Doing this with only natural derivatives appearing, requires, perforce, introducing a constant with the dimensions of time. This is not permitted by our basic axioms, and, indeed, departs from the assumptions underlying the above-mentioned proof (which posits that $\az$ is the only new constant).
\par
One may try, alternatively, to interpolate between the two kinetic Lagrangians, by changing smoothly the order of the time derivative in the kinetic Lagrangian, from $2$ to $1$, going through a sequence of fractional time derivatives, and, thus constructing a nonlocal umbrella theory. This is allowed by the above-mentioned proof. In Sec. \ref{family}, I describe such a sequence, but for reasons explained there, I do not know how to implement, in a sensible way, a continuous interpolation based on such a sequence.
\par
The other potential issue with the linear DML model is the poorer convergence properties, of the expressions for $\psi$, than those of the Newtonian potential.
We saw in Eq. (\ref{thina}) that $\psi$ diverges logarithmically near a thin mass shell of finite surface density $\S$, and radius $R$: $\psi(r)\approx \azg\pi\S\ln\eps$, where $\eps=|r/R-1|$. This is also the behavior on the symmetry axis, at a distance $d$ from a thin disk of radius $R$ and surface density $\S$, where $\psi(d)=\azg\pi\S\ln(d/\sqrt{R^2+d^2})$.
 This is similar to the divergence in Newtonian gravity near an infinitely thin mass wire.
More generally, because of the extra factor of the radius difference in the denominator of expression (\ref{pasafalma}) for $\psi$ (compared with the Newtonian potential), the discontinuities in $\psi$ near discontinuities in the mass distribution are more severe than those of the Newtonian potential.
\par
This makes the model inapplicable to some idealized mass distributions, since it leads to unacceptable phenomenology in such cases. For example, in a spherical system, where there is a finite jump in density at some radius $R$, the dominant behavior
of $\psi(r)$ near $R$ is $\propto |r-R|\ln|r-R|$, so $\grad\psi$ diverges there, logarithmically.
As in Newtonian dynamics, discontinuities of this kind become more severe in thin-disk geometries. For example, $\psi$ diverges everywhere on an infinitely thin disk of finite surface density. For a disk of finite thickness, the thickness enters as a cutoff in the logarithm.
\par
The DML models described below in Secs. \ref{generalizations} and \ref{shell}, have better convergence properties, some even better than those of the Newtonian potential.

\section{Extension to a family of nonlinear DML models modifying gravity and inertia \label{generalizations}}
If we forgo linearity, we can generalize the model described in Sec. \ref{path}, whose Lagrangian is given by Eq. (\ref{genlag}), into a one-parameter family -- all DML models -- underlaid by the Lagrangian
\beq L= \int d^3r \left\{\azg\r[\frac{1}{2\b}(|\vr^{(2/\b)}|^2)^\b-\psi]-q(\b)(\Lap{(2\b-1)/4}\psi)^2\right\}.  \eeqno{genmol}
\par
When $\r$ is made up of point masses, $m_p$, on trajectories $\vr_p(t)$, we have $\r(\vr,t)=\sum_p m_p\d^3[\vr-\r_p(t)]$, so
\beq L= \azg\sum_p m_p[\frac{1}{2\b}(|\vr_p^{(2/\b)}|^2)^\b-\psi(\vr_p)]-q(\b)\int d^3r~(\Lap{(2\b-1)/4}\psi)^2.  \eeqno{genlip}
\par
The kinetic Lagrangian here involves fractional time derivatives -- a nonlocal operation on the trajectory, when the derivative order is not natural.
\par
I give the definition of this fractional derivative that I use, and derive some of its properties, in Appendix \ref{fractional}.
Briefly described, if $\hat\vr(\o)$ is the Fourier transform of $\vr(t)$, i.e.,
\beq \vr(t)=\frac{1}{2\pi}\int_{-\infty}^{\infty}d\o~\hat\vr(\o)e^{i\o t}, \eeqno{gumca}
then, I define
\beq \vr^{(\eta)}\equiv  \frac{1}{2\pi}\int_{-\infty}^{\infty}d\o~|\o|^\eta e^{i[\o t+\eta\frac{\pi}{2}s(\o)]}\hat\vr(\o),  \eeqno{fourin}
where $s(\o)$ is the sign of $\o$ ($s=0$ for $\o=0$).
This definition coincides with the standard derivative for any natural order $\eta=n$, and also ensures that $\vr^{(\eta)}$ is real if $\vr(t)$ is real.
\par
The particular form of the Lagrangian (\ref{genmol}-\ref{genlip}) is arrived at by the following requirements and argumentations:
We want a DML theory, so only $\azg$ can appear as a dimensioned constant. We want the gravitational field equations to still be linear in the density distribution,\footnote{In principle, we could generalize even further to models with a nonlinear gravitational sector, with the two-parameter ($\b$ and $\k$) Lagrangians of the form
\beq L=\int d^3r\{ \azg\r[\frac{1}{2\b}(|\vr^{(\a)}|^2)^\b-\psi]-q(\b,\k) (\Lap{\c}\psi)^\k\},  \eeqno{gensap}
with $\a=2/\b(\k-1)$, $\c=(2\b\k-2\b-1)/2\k$, for dimensional consistency (which also ensures scale invariance). But I do not study further such models.} to facilitate calculation of $\psi$ from $\r$; so, the free $\psi$ Lagrangian density is quadratic in $\psi$, and $\psi$ couples to $\r$ through $\r\psi$. Matter equations of motion are governed by the free matter action. To ensure the universality of free fall in the $\psi$ field, we need the matter kinetic term to also be proportional to $\r$, as the only appearance of mass. This gives rise to the form of the first two terms. From this form follows that $\psi$ has no mass dimensions. The only mass dimensions, besides those of $\r$ appear in $\azg$, which must then appear as it does. I also want $\psi$ to appear only through its derivatives of positive order, so that the theory is invariant to $\psi\rar\psi+{\rm const.}$ (As usual, adding a constant to $\psi$ in the second term adds an immaterial constant $\propto M$ to the Lagrangian.)  $\b$ is our parameter, and the order of the time derivative of $\vr$ and that of the Laplacian in the free $\psi$ term are then determined by $\b$ for consistency of the length and time dimensions. The resulting dimensional consistency, and the sole appearance of the constant $\azg$, ensure scale invariance.
\par
The DML models are not linear, because for $\b\not = 1$ the kinetic Lagrangian density is not quadratic in $\vr$. But the gravitational sector is still linear, and $\psi$ is easily calculated for arbitrary mass distributions.
To ensure invariance to Galilei boosts, under which $\vr^{(\a)}$ is invariant for $\a>1$ (and for $\a=\b=1$), we limit ourselves to $\b<2$.
(The case $\b=2$ is discussed separately below, in Sec. \ref{special}.)
The case $\b=1$ is the linear model of Sec. \ref{path}.
\par
The numerical coefficient $q(\b)$ is chosen to ensure the standard normalization of $\azg$ (and thus of $\az$) for which the mass-asymptotic-speed relation is $V^4_\infty=M\azg$ , which gives (see Sec. \ref{virgen} below)
\beq  q(\b)=\frac{(4-2\b)\Gamma(2-\b)}{4^\b\pi^{3/2}\Gamma(\b-1/2)}. \eeqno{quqara}
Using expression (\ref{appmaca}) for the variation of the free $\psi$ Lagrangian, the field equation for $\psi$ is
\beq \Delta^{[\b-1/2]}\psi=\frac{\azg}{2q(\b)}\r.  \eeqno{latedfy}
\par
Nonlocal, relativistic MOND extensions (involving, e.g., the inverse of the D'Alembertian) were proposed and studied, for example, in Refs. \cite{deffayet14,woodard15a,kim16}.
\par
Using expression (\ref{tarataew}) for the corresponding Green's function, we have (for $\b<2$, as we assume)
\beq  \G_{\b-1/2}(\vr)=-\frac{\Gamma(2-\b)}{4^{\b-1/2}\pi^{3/2}\Gamma(\b-1/2)}|\vr|^{2\b-4}.  \eeqno{natram}
Thus,
\beq  \psi(\vr)=-\frac{\azg}{4-2\b}\int d^3r'\frac{\r(\vr')}{|\vr-\vr'|^{4-2\b}}.  \eeqno{natrlus}
\par
Regarding the EoM of the masses $m_p$, it is hard to vary the kinetic action for general $\b$. The variation of the action, $I=\int dt~L$, under an infinitesimal change $\vr_p(t)\rar \vr_p(t)+\d \vr_p(t)$ is of the form
\beq \d I=m_p\int dt\{\vec\A([\vr_p])-\grad\psi(\vr_p)\}\cdot\d\vr_p(t) +{\rm end~terms~difference}.    \eeqno{mustara}
Here, $\vec\A([\vr_p])$ is some (generally nonlocal) functional of the whole trajectory and of the momentary value of $\vr_p$ (hence the square bracket around $\vr_p$ in the argument).
The particle EoM is thus
\beq  \vec\A[\{\vr_p(t)\}]=\grad\psi(\vr_p), \eeqno{eqayuka}
where, from Eq.(\ref{natrlus}),
\beq  \grad\psi(\vr)=\azg\int d^3r'\frac{\r(\vr')(\vr-\vr')}{|\vr-\vr'|^{6-2\b}}.  \eeqno{natmish}
\par
The dimensions of $\grad\psi$ are $[\ell]^{2\b-1}[t]^{-4}$. Thus, $\vec\A[\{\vr_p(t)\}]$, which has the same dimensions -- and which is constructed from only the trajectory, with no dimensioned constants appearing -- must have the following scaling properties:
\beq   \vec\A[\{\l\vr(t/\z)\}]=\l^{2\b-1}\z^{-4}\vec\A[\{\vr(t)\}].   \eeqno{hyup}
\par
Before discussing further some predictions of these models, we note that they can be free of some of the difficulties with the linear model that I discussed in Sec. \ref{limitations}. Since the particle Lagrangian is time-nonlocal, the models do not necessarily suffer from the Ostrogradsky instabilities. In addition, as seen from Eq. (\ref{natrlus}), the convergence properties of $\psi$ and $\grad\psi$ are better than those of the linear model.

\subsection{Virial relation and rotation curves \label{virgen}}
While it is hard to derive $\vec\A$ from the Lagrangian (\ref{genlip}), we can derive, directly from the action, a useful virial relation, that does not require knowledge of $\A$ as explained in Ref. \cite{milgrom94c}.
\par
 Consider a small change in a trajectory that satisfies the model's EoM, $\d\vr_p(t)=\eps\vr_p(t)$, describing an infinitesimal dilatation of the trajectory. The change in the action under this variation is calculated directly to be
\beq \d I=\eps m_p\int dt[(|\vr_p^{(2/\b)}|^2)^\b-\vr_p\cdot\grad\psi(\vr_p)].  \eeqno{varvare}
On the other hand, $\d I$ is also given by the general expression (\ref{mustara}), where now the first term vanishes because $\vr_p$ solves the EoM. It follows then that $m_p\int dt[(|\vr_p^{(2/\b)}|^2)^\b-\vr_p\cdot\grad\psi(\vr_p)]$ is a difference between two end terms.
Now make the assumption that underlies virial relations, in general, namely that the trajectories are bounded, and thus the difference in end terms is finite. Divide then by the integration time and let it go to infinity. The contribution of the end terms vanishes, and we are left with the virial relation
\beq \av{(|\vr_p^{(2/\b)}|^2)^\b}\_T=\av{\vr_p\cdot\grad\psi(\vr_p)}\_T, \eeqno{vishnar}
where $\av{}\_T$ is the long-time average.\footnote{If the trajectory is periodic, the average can be taken over one period.}
Multiplying by $m_p$, summing over $p$, and using Eq. (\ref{natmish}) for $\grad\psi$ (written for the discrete case), we get
\beq \sum_p m_p\av{(|\vr_p^{(2/\b)}|^2)^\b}\_T=\azg\av{\sum_{q<p}\frac{m_pm_q}{|\vr_p-\vr_q|^{4-2\b}}}\_T.   \eeqno{nitini}
This can be expressed as a generalized $M-\s$, DML relation, by writing it as
\beq \av{\D r}^{4-2\b}\av{|\vr_p^{(2/\b)}|^{2\b}}=M\azg,   \eeqno{jatersa}
where
\beq  \av{\D r}\equiv [\av{\sum_{q<p}\frac{\m_p\m_q}{|\vr_p-\vr_q|^{4-2\b}}}\_T]^{-1/(4-2\b)}  \eeqno{sasar}
is some measure of the mass-weighted average (and time-averaged) intermass distance, and  $\av{|\vr_p^{(2/\b)}|^{2\b}}$ is the mass-weighted average of $\av{(|\vr_p^{(2/\b)}|^2)^\b}\_T$. The left-hand side, which has dimensions of $v^4$, is some measure of the (time-averaged) velocity dispersion of the system ($M$ is the total mass, and $\m_p=m_p/M$).
\subsubsection{Rotation curves}
For a circular orbit of a test particle, in the midplane of an axisymmetric, plane-reflection-symmetric mass distribution -- relevant for rotation curves in such galactic systems as disk galaxies -- the virial relation gives the circular speed
\beq  V^4(r)=r^{5-2\b}|\partial\psi/\partial r|,  \eeqno{karted}
with $\psi$ given by Eq. (\ref{natrlus}).
\par
For a point central mass, or asymptotically far from any bounded mass, $\psi=-(4-2\b)^{-1})M\azg r^{2\b-4}$;
so $V^4(r)=M\azg$, the value of the predicted asymptotic speed in all MOND theories (with the standard normalization of $\az$).
This also justifies the above choice of the coefficient $q(\b)$.
\subsubsection{Two masses on circular orbits}
Two point masses $m_1$, $m_2$ on circular orbits around each other, a distance $R$ apart, orbit a center at distances $r_1$ and $r_2$, respectively ($r_1+r_2=R$), with a common angular frequency $\o$.
The EoMs (\ref{eqayuka}) of the two masses, tell us that $m_1\vec\A[\{\vr_1(t)\}]= m_2\vec\A[\{\vr_2(t)\}]$; so, the scaling property (\ref{hyup}) of $\vec\A$ implies that $r_2/r_1=(m_1/m_2)^{1/(2\b-1)}$.
Substituting in the virial relation, we get that the relative velocity, $V_{12}$ of the two masses, is given by
\beq  V_{12}^4=\o^4R^4=M\azg[\m_1^{1/(2\b-1)}+\m_2^{1/(2\b-1)}]^{2\b-1}.   \eeqno{smatra}
This generalizes expression (\ref{twobo}) for the linear, $\b=1$ case.

\subsection{Special cases  \label{special}}
\subsubsection{$\b=2$}
The limiting model with $\b=2$ is interesting. Our Eqs. (\ref{quqara})-(\ref{natrlus}) seem singular in this limit. This case was the main subject of Ref. \cite{giusti20}, where a regularization procedure was suggested to handle this singularity. But the required result can be obtained directly by taking the limits of our expressions. It is readily seen that in the limit $\b\rar 2$, $q(\b)\rar 1/4\pi^2$.
Expanding expression (\ref{natrlus}) in $2-\b$, we have
\beq \psi=-\frac{\azg}{4-2\b}\int d^3r'~\r(\vr')+\azg\int d^3r'~\r(\vr')\ln|\vr-\vr'|+O(2-\b).  \eeqno{numai}
The first term diverges, but it is the constant $M\azg/(4-2\b)$, and can be ignored by regularization,\footnote{``Regularization'' here means that, for any $\b$, we take $\psi$ to be given by expression (\ref{natrlus}) plus the constant $M\azg/(4-2\b)$.} leaving us in the limit with
\beq \psi=\azg\int d^3r'~\r(\vr')\ln|\vr-\vr'|.  \eeqno{numbook}
Indeed, if we take the limit directly in expression (\ref{natmish}) for $\grad\psi$ there is no divergence, since
\beq \grad\psi\rar \azg\int d^3r'\frac{\r(\vr')(\vr-\vr')}{|\vr-\vr'|^2},  \eeqno{kareta}
as is also gotten from Eq. (\ref{numbook}).
\par
Reference \cite{giusti20}, however, did not use our DML kinetic term, but assumed the standard Newtonian inertial term. His theory is thus not a DML one, but hinges on a length, not an acceleration constant (see further discussion of this in Sec. \ref{path}).
\par
Our kinetic Lagrangian is $\frac{1}{4}\vv^4$, which gives an EoM
\beq  \frac{d(v^2\vv)}{dt}=v^2\va+2(\vv\cdot\va)\vv=\azg\int d^3r'\frac{\r(\vr')(\vr-\vr')}{|\vr-\vr'|^2}. \eeqno{glamat}
The theory is local and quasilinear, in the sense described in the Introduction.
This model is, however, not Galilei invariant.
\par
The kinetic term has the form of the lowest order, special-relativistic correction to Newtonian inertia: $d[\c(v)\vv]/dt$.
\par
The virial relation (\ref{nitini}) reads in this case
\beq  \av{v^4}=\frac{\azg}{M}\sum_{q<p}m_p m_q=\oot M\azg(1-\sum_p\m_p^2),    \eeqno{lireq}
where $M$ is the total system mass, $\av{v^4}=M^{-1}\av{\sum_p m_p v_p^4}\_T$ is the mass-weighted (and long-time) average of $v^4$ (measured in the rest frame of the center of mass of the system), and $\m_p=m_p/M$. This expression of the DML $M-\s$ relation is to be compared with the virial relation for DML systems that holds in a large class of modified gravity theories given above in Eq. (\ref{virimg}).
Here, the definition of $\s$ that enters is the fourth root of the mass-averaged $v^4$, compared with the square root of the mass-averaged $v^2$. The dependence on the mass ratios is also different.

\subsubsection{$\b=3/2$  \label{betihal}}
For $\b=3/2$, we have $\psi=\az\fN$, where $\fN$ is the Newtonian potential.
Then, the usual shell theorems apply. This is a pure modified-inertia model. The predicted rotation curve, from Eq. (\ref{karted}), implies
$a^2=\az\gN$, in line with the general theorem for modified-inertia MOND theories \cite{milgrom94a}.
The velocity difference for two masses on circular orbits around each other is, likewise, what is predicted in purely modified-inertia theories, as given in Eq. (\ref{circulla}).

\section{Families of DML models generalizing the AQUAL and QUMOND DMLs \label{shell}}
Another interesting one-parameter ($\z$) family of DML models is underlaid by the Lagrangian density
\beq \L= \azg\r[\frac{1}{2\b}(|\vr^{(\a)}|^2)^\b-\psi]-\frac{1}{8\pi\z}[(\grad\psi)^2]^\z,  \eeqno{getrap}
with $\b=(2\z+1)/(4\z-2)$, $\a=(2\b-1)/\b$.
\par
The field equation of $\psi$ is
\beq \div\{[(\grad\psi)^2]^{\z-1}\grad\psi\}=4\pi\azg\r.  \eeqno{lalater}
For $\z=3/2$, we get $\a=\b=1$, and {\it the model is the DML of AQUAL}. This is the maximum value of $\z$ that gives $\a>1$ or $\a=\b=1$, which are sufficient for Galilei invariance.
For $\z=1$, we have $\b=3/2$, and $\a=4/3$, which is the special DML model, with $\b=3/2$, in the family of Sec. \ref{generalizations}, discussed in Sec. \ref{betihal}.
\par
While both the inertial and the gravitational equations for these models are nonlinear, their advantage is that they do satisfy the usual shell theorems -- in spherical systems, $\psi'(r)$ depends only on the total mass within $r$ -- as shown in Sec. \ref{sphesi} below.

\subsection{Virial theorem}
Here too, it is hard to derive the particle EoM, but as before, we can derive a global virial relation for the trajectory of a particle in a given gravitational field.
As in the derivation of the particle virial relation (\ref{vishnar}), here we have
\beq \av{(|\vr_p^{(\a)}|^2)^\b}\_T=\av{\vr_p\cdot\grad\psi(\vr_p)}\_T. \eeqno{vishpal}

\subsection{Spherical systems \label{sphesi}}
Applying Gauss's theorem to the field equation (\ref{lalater}), we get that
\beq |\psi'(r)|=\left[\frac{M(r)\azg}{r^2}\right]^{1/(2\z-1)}=[\az\gN(r)]^{1/(2\z-1)},  \eeqno{sfufa}
where $M(r)$ is the mass enclosed within $r$.

\subsection{Rotation curves}
For a circular trajectory in the midplane of an axisymmetric, plane-symmetric configuration we get from the virial relation,
given that $\av{(|\vr_p^{(\a)}|^2)^\b}\_T=\o^{2\a\b}r^{2\b}$, where $r$ is the orbital radius, and $\o$ the frequency.
\beq V^4(r)=r^2\left| \frac{\partial\psi}{\partial r} \right|^{2\z-1}.  \eeqno{hatreq}
For circular orbits in a spherical system, we get from Eq. (\ref{sfufa})
\beq  V^4(r)=M(r)\azg.   \eeqno{vitagy}
Thus, for circular orbits in a spherical field we get the same relation between acceleration and Newtonian acceleration, $a(r)=(\az\gN)^{1/2}$ as in the DML of pure modified gravity theories. But this does not hold for other trajectories, or in nonspherical mass distributions.
\subsection{Analogous family that extends the QUMOND DML}
In an analogous way, we can construct a family of DML models of which the DML of QUMOND is a member.
The Lagrangian density of this family is
\beq \L= \azg\r[\frac{1}{2\b}(|\vr^{(\a)}|^2)^\b-\psi]+\frac{1}{8\pi\z}[(\grad\tilde\f\_N)^2]^\z-\frac{1}{4\pi}\grad\psi\cdot\grad\tilde\f\_N,  \eeqno{getrap}
with $\b=(4\z-1)/2$, $\a=(2\b-1)/\b$.
\par
The field equation for $\tilde\f\_N$ equates it with $\az\fN$, where $\fN$ is the Newtonian potential, and
the field equation of $\psi$ is
\beq \D\psi=\div\{[(\grad\tilde\f\_N)^2]^{\z-1}\grad\tilde\f\_N\}.  \eeqno{qudam}
The DML of QUMOND is gotten for $\z=3/4$ ($\a=\b=1$). This is also the minimal value of $\z$ that gives $\a>1$ or $\a=\b= 1$, which is sufficient to ensure Galilei invariance. We see that $\a<2$ for any finite $\z$.
\par
Like QUMOND itself, these models require solving only the (linear) Poisson equation, twice, with an algebraic step in between.
\par
I do not further discuss this family, which yields results similar to those above.

\section{Family of one-parameter models bridging the linear DML model and Newtonian dynamics  \label{family}}
The Newtonian EoM (\ref{newt}) and the linear, DML model  (\ref{pasafma})-(\ref{lag}) can be written in a single parametrized form -- here for the discrete-masses case -- as
 \beq (-1)^\eta\vr_p^{(2\eta)}=G\az^{\eta-1}\sum_{q\not = p}\frac{m_q(\vr_p-\vr_q)}{|\vr_p-\vr_q|^{\eta+2}},  \eeqno{pasafra}
 derived from the action
\beq I_\eta=\int~ dt~L_\eta;~~~~~~~L_\eta=\sum_p \oot m_p[\vr_p^{(\eta)}]^2+\frac{G\az^{\eta-1}}{\eta}\sum_{q<p}\frac{m_p m_q}{|\vr_p-\vr_q|^\eta}.    \eeqno{lageta}
For $\eta=1$  we get the former, and for $\eta=2$ we get the latter.
\par
I wish to define a family of theories, spanned by the values of $1\le\eta\le 2$, that coincide with the above two theories for the end values.
These have the kinetic term, and dependence on positions and masses, as given by expression (\ref{lageta}). As to the prefactor of the gravitational term, dimensional consideration, and the fact that it has to be constructed with $G$ and $\az$, dictate that it is $G\az^{\eta-1}$ up to a dimensionless factor $\xi(\eta)$. We thus get a family of theories governed by the Lagrangians
\beq L_\eta=\sum_p \oot m_p[\vr_p^{(\eta)}]^2+\xi(\eta)\frac{G\az^{\eta-1}}{\eta}\sum_{q<p}\frac{m_p m_q}{|\vr_p-\vr_q|^\eta},    \eeqno{lagbus}
with $\xi(1)=\xi(2)=1$.
\par
We can construct such a family in the framework where $\psi$ is a degree of freedom. To this end, I first notice that the Poissonian Lagrangian $\propto \int d^3r(\grad\psi)^2$ can be written in terms of a fractional Laplacian, since, as shown in Appendix A, Sec. \ref{freelag},
\beq  \int d^3r(\grad\psi)^2=\int d^3r(\Lap{1/2}\psi)^2.  \eeqno{jajaot}
Thus, the Lagrangian underlying Newtonian gravitational dynamics can be written as
\beq L_\eta= \sum_p \oot m_p[\vr_p^{(\eta)}]^2-\int d^3r~\r\psi-\frac{1}{8\pi G\az^{\eta-1}}\int d^3r(\Lap{1/2}\psi)^2,  \eeqno{genlagla}
where $\r(\vr,t)=\sum_p m_p\d^3[\vr-\vr_p(t)]$, and
with $\eta=1$. This is to be compared with the Lagrangian for the DML theory Eq. (\ref{genlag}), which can be written in an analogous form
\beq L_\eta= \sum_p \oot m_p[\vr_p^{(\eta)}]^2-\int d^3r~\r\psi-\frac{1}{2\pi^2 G\az^{\eta-1}}\int d^3r(\Lap{1/4}\psi)^2,  \eeqno{genlop}
with $\eta=2$.
\par
How is this Lagrangian to be defined for other $\eta$ values?
Start with the generalization of the gravitational term, where, to ensure linearity of the theory we
take it as $(\Lap{\l}\psi)^2$, with $\l(\eta)$ some function of $\eta$ satisfying $\l(1)=1/2$, $\l(2)=1/4$.
Dimensional consistency dictates that the gravitational prefactor has to be $G^{-1}\az^{1-\eta}$, and that $\l(\eta)=(3-\eta)/4$.
The desired family of theories is thus underlaid by the Lagrangians
\beq L_\eta= \sum_p \oot m_p[\vr_p^{(\eta)}]^2-\int d^3r~\r\psi-\frac{1}{2\chi(\eta) G\az^{\eta-1}}\int d^3r(\Lap{\l(\eta)}\psi)^2,  \eeqno{regenlag}
where $\chi(1)=4\pi$, and $\chi(2)=\pi^2$.
\par
By Eq. (\ref{appmaca}), which describes the variation of the gravitational term with $\psi$, the field equation for $\psi$ is
\beq \Lap{2\l(\eta)}\psi=\chi(\eta)G\az^{\eta-1}\r.   \eeqno{filequr}
\par
By Eq. (\ref{tarataew}), the Green's function of $\Lap{2\l(\eta)}$ is
\beq  \G_{2\l}=\frac{-\Gamma(\eta/2)}{4^{(3-\eta)/2}\pi^{3/2}\Gamma(3/2-\eta/2)}\frac{1}{|\vr|^\eta}\equiv -\frac{\t(\eta)}{|\vr|^\eta}.  \eeqno{gregen}
The field $\psi$ is then given by
\beq  \psi(\vr)=-\t(\eta)\chi(\eta)G\az^{\eta-1}\int d^3r'\frac{\r(\vr')}{|\vr-\vr'|^\eta}. \eeqno{ratade}
Thus, we get the theory described by the Lagrangian (\ref{lagbus}) with $\xi(\eta)=\eta\t(\eta)\chi(\eta)$.
Since we are free to choose $\chi(\eta)$ under the ends constraints -- remembering that each choice yields a different family of models -- one simplifying choice is $\xi(\eta)=1$ [$\chi=1/\eta\t(\eta)$], which satisfies the constraints, since $\t(1)=1/4\pi$, and $\t(2)=1/2\pi^2$. This would give a family of theories with
\beq  \psi(\vr)=-\frac{G\az^{\eta-1}}{\eta}\int d^3r'\frac{\r(\vr')}{|\vr-\vr'|^\eta} \eeqno{ratpaser}
for all $\eta$ values.\footnote{Note that, using our definition of fractional time derivatives, Eq. (\ref{fourapi}), a kinetic Lagrangian density $\vr^{(\z)}\cdot\vr^{(\xi)}$, with $\z\not =\xi$ does not generalize the one above, since it can be seen that $\int dt~\vr^{(\z)}\cdot\vr^{(\xi)} =\cos{[\pi(\z-\xi)/2]}\int dt[\vr^{(\eta)}]^2$, with $\eta=(\z+\xi)/2$; so it gives the same theory (after absorbing the cosine factor into $\azg$), provided $|\z-\xi|$ is not an odd integer, in which case the theory is degenerate.}
\par
Consider now the kinetic, particle Lagrangian. Its change under a small variation, $\d\vr(t)$, of $\vr(t)$ is derived in Sec. \ref{varparac}, where I find that
\beq \d\int_{-\infty}^\infty dt~[\vr^{(\eta)}(t)]^2=2\int_{-\infty}^\infty dt~\vr^{[2\eta]}(t)\cdot\d\vr(t),  \eeqno{haters}
where
\beq  \vr^{[2\eta]}\equiv \frac{1}{2\pi}\int_{-\infty}^{\infty}d\o~|\o|^{2\eta}\hat\vr(\o) e^{i\o t}.    \eeqno{lamamain}
This leads to the particle EoM
\beq \vr_p^{[2\eta]}(t)=\grad\psi(\vr_p).  \eeqno{eomeom}
In the discrete, point-masses description
\beq \grad\psi(\vr_p)=G\az^{\eta-1}\sum_{q\not =p}\frac{m_q(\vr_p-\vr_q)}{|\vr_p-\vr_q|^{\eta+2}}.   \eeqno{mioreq}
Equation (\ref{eomeom}) can be written as
\beq  \frac{d\vP_p}{dt}=m_p\grad\psi(\vr_p),   \eeqno{mushtar}
where
\beq  \vP_p=\frac{m_p}{2\pi}\int_{-\infty}^{\infty}d\o~(i\o)^{-1}|\o|^{2\eta}\hat\vr_p(\o) e^{i\o t}.    \eeqno{lamatar}
So, we identify $\vP_p$ as the particle momentum. It is a real vector, a time-dependent, nonlocal functional of the trajectory, such that, for an isolated many-particle system, the total momentum $\vP=\sum_p\vP_p$ is conserved.
\par
We can also define a (nonlocal) kinetic energy, $E_{Kp}$, that satisfies $dE_{Kp}/dt=-m_p\vv_p\cdot\grad\psi(\vr_p)=-d[m_p\psi(\vr_p)]/dt$, such that $E_{Kp}+m_p\psi(\vr_p)$ is conserved.
\par
Note that $\vr^{[2\eta]}$ does not generally equal the derivative $\vr^{(2\eta)}$: By Eq. (\ref{fourin})
we have
\beq  \vr^{(2\eta)}= \frac{1}{2\pi}\int_{-\infty}^{\infty}d\o~|\o|^{2\eta}\hat\vr(\o) e^{i[\o t+\eta\pi s(\o)]},    \eeqno{lamaru}
to be contrasted with expression (\ref{lamamain}). For the case of local theories, with $\eta=n$ a natural number, we have
\beq \vr^{[2n]}=(-1)^n\vr^{(2n)}, \eeqno{nukva}
which aligns with the signs in the Euler-Lagrange equation for such theories. In particular, for the two limiting cases $\eta=1,2$, Eq. (\ref{eomeom}) gives the corresponding EoMs discussed above.
\par
Another difference between the two quantities is that under time reversal, $\vr(t)\rar\vr(-t)$, which is tantamount to $\hat\vr(\o)\rar \hat\vr(-\o)$,
 also $\vr^{[2\eta]}(t)\rar\vr^{[2\eta]}(-t)$, from its definition.
In contrast, $\vr^{(2\eta)}(t)\rar (-1)^{2\eta}\vr^{(2\eta)}(-t)$ for natural values of $2\eta$, and there is no definite sign for other values.
\par
Thus, importantly, with $\vr^{[2\eta]}$ as the kinetic term, the EoM (\ref{eomeom}) is time-reversal invariant (even for odd values of $2\eta$)\footnote{We see from expression \ref{sateyu}, of the action in terms of the Fourier components, that the action is manifestly time-reversal symmetric.} [which it would not be with $\vr^{(2\eta)}$].
\par
This family of models are also Galilei invariant, because under a Galilei transformation $\hat\vr(\o)\rar\hat\vr(\o)+\vv_0\d'(\o)$, and
$|\o|^{2\eta}\d'(\o)=0$ for $\eta> 1/2$; so, $\vr^{[2\eta]}$ is invariant\footnote{This can be seen already at the level of the action, where under a Galilei boost, the change in the action is $\propto \int d\o |\o|^{2\eta}\{\d'(\o)\vv_0\cdot[\hat\vr(\o)-\hat\vr(-\o)]+\vv_0^2[\d'(\o)]^2\}$. The last term is independent of $\hat\vr(\o)$, i.e., is an immaterial constant, and the first term vanishes for $\eta>1/2$.}
\par
All the theories, with $1\le\eta\le 2$, are linear; only the end ones -- Newton's and the DML -- are local; the rest are, clearly, time nonlocal as the fractional derivative requires knowledge of the full trajectory. And, only the DML model is scale invariant.

\par
As another instance of using fractional time derivatives in connection with MOND, I note Ref. \cite{barrientos21}, which discussed modification to the Friedmann cosmological equations replacing the standard, general-relativistic time derivatives of the cosmological scale factor by fractional ones left as free parameters -- with suggested connection to MOND cosmology.
\par
It would be interesting to connect the strict DML model discussed in Sec. \ref{path} with Newtonian dynamics under some umbrella theory that have them as low-, and high-acceleration limits, respectively.
Can we do this based on the above family spanned by $\eta$, and that covers the full range by making $\eta$ a function of $a/\az$, with $a$ some acceleration attribute? I do not know how to do this in a sensible way, since different constituents of the system have different accelerations -- some may be higher, some lower than $\az$. So, while it is easy to assign an $\eta$ value to the left-hand side of the EoM (\ref{eomeom}), it is not clear how to do it for the right-hand side.

\section{Summary -- some lessons learned  \label{lessons}}
All nonrelativistic MOND theories, or models, that were studied to date modify either only the gravitational sector (i.e., the Poisson equation), or only inertia (i.e., the free particle action). The models presented here show that it may be productive, or even necessary, to modify both. This is expected on general grounds if we take the path of modified inertia, because the Einstein-Hilbert action, which reduces to the Poisson action in the nonrelativistic limit, may be thought of as the inertial actions of the gravitational field.
\par
Another lesson that pertains to theories that modify Newtonian inertia is that since the inertia term in the particle EoM is no longer the acceleration, such theories do not predict an acceleration field, as is the case in Newtonian dynamics and MOND modified gravity theories. In pure modified-inertia theories, such as those discussed in Ref. \cite{milgrom23}, the gravitational field does enter as an acceleration (the gradient of the potential), but the EoM equates it with some more involved functional of the trajectory than the acceleration itself (this is also true in special relativity, which modified Newtonian inertia).
\par
This is crucial in the context of galactic dynamics, where, by and large, we do not even measure accelerations directly, let alone higher time derivative, or full trajectories.\footnote{Except in special cases such as for ideal circular orbits in an axisymmetric field, where the trajectory is fully specified by the measure orbital radius, and the velocity at one point.} At best, we combine measured velocities and radii to determine, or estimate, accelerations of certain test particles, at certain positions, and compare them with the calculated gravitational accelerations.
For example, in using rotation curves of disk galaxies we measure the accelerations of test particles on circular orbits, in the plane of the disk.
If a modified-inertia theory underlies the dynamics, this procedure can lead us astray, because the predicted kinematic acceleration at a given position depends not only on the position, but also on more details of the trajectory of the test particles used to map the field.
\par
The models we studied here demonstrate that in theories that modify both inertia and gravity, even gravity itself does not enter as an acceleration field {\it in the DML}, which makes the acceleration itself an even less relevant quantity in the DML.
\par
This may sound paradoxical in MOND, which is widely advertised as hinging on accelerations. MOND certainly does hinge on accelerations, inasmuch as we are speaking of transitional Newtonian-to-DML phenomena, where $\az$ itself plays the crucial role. However, the constant that is relevant in the DML -- ``the second MOND constant'', $\azg$, which is perhaps the more fundamental of the two (see discussion in Ref. \cite{milgrom15}) -- is not an acceleration, and there is no preferred acceleration scale in the DML. The often-encountered DML relation $g\approx(\gN\az)^{1/2}$ should be read as $g\approx(M\azg)^{1/2}/R$. The quantity $MG/R^2$, which we call the Newtonian gravitational acceleration, because it is that in the Newtonian regime, is undefined and has no meaning in a DML-governed world, as the above pure-DML models demonstrate.\footnote{Sometimes, we find it convenient to use our acquaintance with the Newtonian regime, and the knowledge of its constant $G$, to express DML results in therm of the familiar -- but DML irrelevant -- $\gN$, by introducing the constant $\azg/G$, which we call $\az$.}
\par
We do, as much as we can, express results in terms of accelerations, because accelerations are what we can measure at present in galactic dynamics, even if acceleration is not the most directly relevant quantity.
\par
Another important lesson, learned already to an extent, e.g., from studies of some pure modified-inertia models \cite{milgrom23}, concerns the EFE.
In Newtonian dynamics, the internal dynamics of a small system, falling freely in the gravitational field of a large mother system, are oblivious to this external motion (barring some tidal effects if the external field is not constant across the subsystem). In MOND, with its generic nonlinearity, the internal dynamics can be greatly affected by the external free fall \cite{milgrom83,bm84}. In the modified-gravity studies to date, this effect depends in a rather concrete manner on the momentary acceleration of the external motion. In some of the modified-inertia models discussed in Ref. \cite{milgrom23} the effect depends also on the frequencies of the external and internal motions.
\par
The DML models presented in Secs. \ref{generalizations} and \ref{shell} exhibit an even larger variety of EFE workings. For example, the linear theory we started with predicts no EFE. In other models of all the DML the families, the EFE may depend on the external orbit in more complicated ways. In these models, accelerations do not even appear as fundamental orbital attributes. What enters, and may determine the workings of the EFE, are quantities with the dimensions of some fractional time derivative of the trajectory, which is not even a local attribute, but depends on the full trajectory. To recapitulate, {\it the EFE may have nothing to do with internal and external accelerations. In the DML there is no special role to accelerations.}
\par
Finally, an important, but perhaps little known, fact is that some primary DML predictions of MOND make use not only of the DML axioms, but also rely crucially on the relatively fast transition to the Newtonian regime as we cross $\az$ from low to high accelerations. This reliance on the vicinity of Newtonian behaviors to the DML is evident e.g., in the derivations in Refs. \cite{milgrom14a,milgrom24}. A case in point that I use to explain the issue is the $M-\s$ relation, derived, e.g., in Ref. \cite{milgrom14a}, which states that with an appropriately defined system velocity dispersion, $\s$, and representative mass, $M$, the quantity $\chi=\s^4/M\azg$ is of order unity for all isolated, stationary (time-independent) DML systems.
\par
Such would have been the case had the characteristic radius, $R$, of any DML system been equal to its MOND radius: $R=\rM\equiv (MG/\az)^{1/2}$, and if the systems satisfied a Newtonian virial relation for some $\s$: $\s^2\approx MG/R$. Neither is the case, however, since DML systems have, by definition, $R\gg\rM$, and yet this argument for the approximate universality of $\chi$ is not fortuitous, as I now show.
\par
The DML axioms -- scale invariance, and the sole appearance of $\az$ and $G$ -- imply that within any two-parameter family of similar, time-independent mass distributions, $\r(\vr)=a\bar\r(\vr/b)$, for any $a,b>0$, $\chi$ takes the same value for any definition of $\s$ and $M$ consistent with the scaling. This follows from the scaling relations discussed in the beginning of Sec. \ref{predictions} (taking there $a=\l\k^{-4}$ and $b=\l$, and imposing time independence). For example, all DML systems with exponential density, $\r(r)=\r_0{\rm e}^{-r/h}$, have the same value of $\chi$, if $\s$ is the velocity dispersion at $5h$ and $M$ is the mass contained within $2h$, irrespective of $\r_0$ and $h$. But, if $\s$ and $M$ are defined arbitrarily, the value of $\chi$ may depend strongly on the exact density distribution.
However, we are after a universal, DML correlation for which $\chi$ is of order unity irrespective of system type; for this $\s$ and $M$ have to be chosen discriminately.
\par
Consider a two-parameter family of isolated, self-gravitating systems as described above, and take $M$ to be the total mass. Choose a definition of a characteristic mean velocity dispersion, $\s$, and a characteristic radius, $R$, satisfying the following two requirements: (a) Newtonian dynamics would justify the virial relation $\s^2\approx GM/R$, between $\s$ and $R$; (b) The ratio $R/\rM$ is a faithful measure of the mean acceleration in the system in the MOND sense; namely, for $R\gg\rM$, DML dynamics apply, and for $R\ll\rM$, Newtonian dynamics do.
The requirement from a MOND theory that Newtonian behavior lies near the DML behavior is crucial, as it implies that members of the family that lie on the $R=\rM$ line in the $M-R$ plane, live approximately in the two worlds at once:
They have approximately the universal, DML $\chi$ value of their family, and they also satisfy approximately the Newtonian virial relation, yielding $\chi\sim 1$.
\par
All this is relevant in the present context. Because the DML models discussed in Secs. \ref{generalizations} and \ref{shell} do not have a proper Newtonian continuation, it is not clear that they predict a distribution of $\chi$ values around unity. Indeed, the coefficient in the $M-\s$ relation (\ref{mavel}) that we derived for the linear, DML model, is not guaranteed to be of order unity for all systems, and it is not clear that the definitions of $\s$ and of the characteristic radius used in the derivation satisfy the above two conditions. If they, indeed, do not, we shall learn from it that the proximity of the Newtonian limit to the DML is, indeed, necessary to make even pure DML primary predictions.

\appendix
\section{Fractional derivatives and their manipulation \label{fractional}}
In this paper, I employ derivatives of fractional order, both space derivatives, appearing here as fractional Laplacians, and fractional time derivatives.
There is much literature on the subject. Some more recent detailed discussions can be found in Refs.
\cite{caputo67,hilfer00,herrmann14,kwasnicki17,garofalo17,stinga19,giusti20}.
\par
There are many ways to define fractional derivatives (of non-natural order). Since derivation of any natural order is an algebraic manipulation in Fourier space -- multiplication by the appropriate power of the Fourier variable -- a natural way to define fractional derivatives, which I use in this paper, is to define them by their action in Fourier space.
I now describe the definition I use and some manipulations and formulas involving such derivatives.
\subsection{Fractional Laplacian}
If $\hat\psi(\vk)$ is the Fourier transform of $\psi(\vr)$; namely,
\beq \psi(\vr)=\frac{1}{(2\pi)^3}\int d^3k~ \hat\psi(\vk){\rm e}^{i\vk\cdot\vr} ,  \eeqno{foutranas}
then, the Fourier transform of $\D\psi(\vr)$ is $-|\vk|^2\hat\psi(\vk)$. We would then tentatively want to define a generalization to a fractional degree $\a$, designated $\Lap{\a}$, such that the Fourier transform, $\widehat{\Lap{\a}\psi}(\vk)$ scales as $(-|\vk|^2)^\a\hat\psi(\vk)=(-1)^\a|\vk|^{2\a}$. This would indeed coincide with the transform of $\D^n\psi(\vr)$ for natural $\a=n$.
However, I also require that if $\psi$ is a real function, so is $\Lap{\a}\psi$. Its transform thus has to satisfy $\widehat{\Lap{\a}\psi}(-\vk)=\widehat{\Lap{\a}\psi}^*(\vk)$.
Writing $(-|\vk|^2)^\a$ as ${\rm e}^{i\xi}|\vk|^{2\a}$, with $\xi$ some phase that does not depend on $\vk$, the reality requirement constrains the phase factor to be real; i.e., either $1$ or $-1$. I also require that $\Lap{1}=\D$; so, the phase factor has to be taken as -1.
I thus define
\beq \Lap{\a}\psi(\vr)=-\frac{1}{(2\pi)^3}\int d^3k~|\vk|^{2\a}\hat\psi(\vk){\rm e}^{i\vk\cdot\vr}d^3k.  \eeqno{foutra}
This definition is in line with those discussed, e.g., in Refs. \cite{kwasnicki17,giusti20}, arrived at based on other justifications. Expression (\ref{foutra}) equals what is defined in those references as $-(-\D)^\a$.
\par
Note that $\Lap{n}=(-1)^{n-1}\D^n$; so, even for natural degrees it coincides with the standard derivatives only for odd $n$, but has the opposite sign for even $n$. Note also that $\Lap{\a}c=0$, for a constant $c$, for any $\a>0$.
\par
I now derive some results that are used in the body of the paper.
\subsubsection{Variation of the free $\psi$ Lagrangian}
In Secs. \ref{generalizations} and \ref{family}, we needed the variation of $\int d^3r~(\Lap\a\psi)^2$, appearing in the free $\psi$ Lagrangian, to first order in a small change $\d\psi$ in $\psi$.
Use the definition (\ref{foutra}) above to write
\beq \int d^3r~(\Lap\a\psi)^2=\frac{1}{(2\pi)^6}\int d^3r\int\int d^3k~d^3k'~(|k||k'|)^{2\a}\hat\psi(\vk){\rm e}^{i\vk\cdot\vr} \hat\psi(\vk'){\rm e}^{i\vk'\cdot\vr}=\frac{1}{(2\pi)^3}\int d^3k~|k|^{4\a}\hat\psi(\vk)\hat\psi(-\vk),  \eeqno{asar}
where the $\vr$ integration was performed to give $(2\pi)^3\d^3(\vk+\vk')$.
Thus,
\beq \d\int d^3r~(\Lap\a\psi)^2=\frac{1}{(2\pi)^3}\int d^3k~|k|^{4\a}[\hat\psi(\vk)\d\hat\psi(-\vk)+\d\hat\psi(\vk)\hat\psi(-\vk)]=\frac{2}{(2\pi)^3}\int d^3k~|k|^{4\a}\hat\psi(\vk)\d\hat\psi(-\vk),  \eeqno{asaras}
since the two terms in the integrand contribute equally.
Comparing with
\beq  \int d^3r~(\Lap{2\a}\psi)\d\psi=-\frac{1}{(2\pi)^6}\int d^3r\int\int d^3k~d^3k'~|k|^{4\a}\hat\psi(\vk){\rm e}^{i\vk\cdot\vr} \d\hat\psi(\vk'){\rm e}^{i\vk'\cdot\vr}=-\frac{1}{(2\pi)^3}\int d^3k~|k|^{4\a}\hat\psi(\vk)\d\hat\psi(-\vk),   \eeqno{jatur}
we see that
\beq \d\int d^3r~(\Lap\a\psi)^2=-2\int d^3r~(\Lap{2\a}\psi)\d\psi.  \eeqno{appmaca}
\subsubsection{The free $\psi$ Lagrangian for $\a=1/2$ equals the free Poisson Lagrangian \label{freelag}}
In Sec. \ref{family}, we needed the result that for $\a=1/2$, the free $\psi$ Lagrangian is the same as the Poisson Lagrangian. Write
\beq \int d^3r~(\Lap{1/2}\psi)^2= \frac{1}{(2\pi)^6}\int d^3r\int\int d^3k~d^3k'~(|k||k'|)\hat\psi(\vk){\rm e}^{i\vk\cdot\vr} \hat\psi(\vk'){\rm e}^{i\vk'\cdot\vr}=\frac{1}{(2\pi)^3}\int d^3k~|k|^2\hat\psi(\vk)\hat\psi(-\vk), \eeqno{jajmit}
and compare with
\beq  \int d^3r~(\grad\psi)^2= \frac{1}{(2\pi)^6}\int d^3r\int\int d^3k~d^3k'~i\vk\cdot i\vk'\hat\psi(\vk){\rm e}^{i\vk\cdot\vr} \hat\psi(\vk'){\rm e}^{i\vk'\cdot\vr}=\frac{1}{(2\pi)^3}\int d^3k~|k|^2\hat\psi(\vk)\hat\psi(-\vk).   \eeqno{matuger}
So the two expressions are equal.
\subsubsection{Green's function of the fractional Laplacian}
We seek $\G_\a(\vr)$ such that
\beq \Lap{\a}\G_\a(\vr)=\d^3(\vr)=\frac{1}{(2\pi)^3}\int d^3k~{\rm e}^{i\vk\cdot\vr}.   \eeqno{miolut}
Comparing with the definition in Eq. (\ref{foutra}) we see that the Fourier transform of $\G_\a$ is
\beq \hat\G_\a(\vk)=-|k|^{-2\a}.  \eeqno{katarew}
Using tables for inverse Fourier transforms, we get
\beq  \G_\a(\vr)=-\frac{\Gamma(3/2-\a)}{4^\a\pi^{3/2}\Gamma(\a)}|\vr|^{-(3-2\a)},\eeqno{tarataew}
where $\Gamma$ is the Gamma function. For the Laplacian itself, $\a=1$, we get the standard $\G_1=-1/4\pi|\vr|$. For our linear DML model ($\a=1/2$), we get $\G_{1/2}=-1/2\pi^2|\vr|^2$.

\subsection{Fractional time derivatives}
Decompose the trajectory $\vr(t)$ into its Fourier components
\beq \vr(t)=\frac{1}{2\pi}\int_{-\infty}^{\infty}d\o~\hat\vr(\o)e^{i\o t} \eeqno{gumcaap}
[since $\vr(t)$ is real, $\hat\vr(-\o)=\hat\vr^*(\o)$].
For natural values of the order $n$, we have
\beq \vr^{(n)}\equiv  \frac{1}{2\pi}\int_{-\infty}^{\infty}d\o~(i\o)^n\hat\vr(\o) e^{i\o t}.  \eeqno{four}
One then takes this to be the definition of the time derivative of any order, $\eta$.
But, we need to specify the definition of $(i\o)^\eta$, which depends on the choice of the logarithm branch in the complex plane. Write
\beq u(\o)\equiv (i\o)^\eta =e^{\eta\ln(i\o)}= e^{\eta[\ln{|\o|}+i\frac{\pi}{2}s(\o)+2i\pi k ]}, \eeqno{defafa}
where $s(\o)$ is the sign of $\o$ ($s=0$ for $\o=0$), and
with the integers $k$ corresponding to the branches of the logarithm.
We want to define the $\eta$ derivative of a real function so that it is also real. This requires
that $u(-\o)=u^*(\o)$, leading to the choice $k=0$
\beq u(\o)= e^{\eta[\ln{|\o|}+i\frac{\pi}{2}s(\o)]}=|\o|^\eta e^{i\eta\frac{\pi}{2}s(\o)}, \eeqno{defaji}
with $|\o|^\eta$ the real, non-negative value.
We thus define
\beq \vr^{(\eta)}\equiv  \frac{1}{2\pi}\int_{-\infty}^{\infty}d\o~|\o|^\eta e^{i[\o t+\eta\frac{\pi}{2}s(\o)]}\hat\vr(\o).  \eeqno{fourapi}
With this definition we also have the usual relation $d\vr^{(\eta)}/dt=\vr^{(\eta+1)}$, and, more generally, $[\vr^{(\eta)}]^{(\z)}=\vr^{(\eta+\z)}$.
\subsubsection{Variation of the particle action  \label{varparac}}
Using definition (\ref{fourapi}), the particle action defined in Sec. \ref{family} is
\beq  I\_P=\oot\int dt~[\vr^{(\eta)}(t)]^2 =\frac{1}{8\pi^2}\int dt\int d\o\int d\o' |\o|^\eta |\o'|^\eta \hat\vr(\o)\hat\vr(\o'){\rm e}^{i\{(\o+\o')t+\eta\frac{\pi}{2}[s(\o)+s(\o')]\}}.  \eeqno{saterq}
The time integration yields $2\pi\d(\o+\o')$; so, we have
\beq  I\_P=\frac{1}{4\pi}\int d\o~|\o|^{2\eta} \hat\vr(\o)\hat\vr(-\o).  \eeqno{sateyu}
Varying over $\vr$ we have
\beq  \d I\_P=\frac{1}{4\pi}\int d\o~|\o|^{2\eta} [\hat\vr(\o)\d\hat\vr(-\o)
+\hat\vr(-\o)\d\hat\vr(\o)]
=\frac{1}{2\pi}\int d\o~|\o|^{2\eta} \hat\vr(\o)\d\hat\vr(-\o),  \eeqno{gagach}
which equals
\beq  \d I\_P=\int dt~\vr^{[2\eta]}\cdot\d\vr(t),  \eeqno{geroch}
where
\beq  \vr^{[2\eta]}= \frac{1}{2\pi}\int_{-\infty}^{\infty}d\o~|\o|^{2\eta}\hat\vr(\o) e^{i\o t}.    \eeqno{lamama}
(Everywhere, $|\o|^{2\eta}$ is the real non-negative value.)

%\clearpage
\end{document}